\documentclass{article}
\usepackage[utf8]{inputenc}
\usepackage{authblk}
\usepackage{amsmath}
\usepackage{amssymb}
\usepackage{verbatim}
\usepackage{amsthm,amsfonts,bm}
\usepackage{MnSymbol}
\usepackage{cite}
\usepackage{graphicx}
\usepackage{color}
\usepackage[colorlinks=true, allcolors=blue]{hyperref}
\usepackage{dsfont}
\usepackage{physics}
\usepackage{indentfirst}
\usepackage{braket}
\usepackage{cancel}
\usepackage{soul}

\definecolor{Red}{rgb}{0.9,0,0}
\definecolor{black}{rgb}{0.0, 0.0, 0.0}

\newcommand{\te}{\theta}

\newcommand{\arcangleru}{%
	\mathord{<\mspace{-10.5mu}\mathrel{\rcurvearrowup}\mspace{2mu}}%
}

\title{One Hundred Years Later: Stern-Gerlach Experiment and Dimension Witnesses}
\author[1]{R. Grossi}
\author[1]{Lucas L. Brugger}
\author[1]{B. F. Rizzuti}
\author[2,3,4]{C. Duarte}

\affil[1]{Departamento de F\'isica, Universidade Federal de Juiz de Fora, MG, Brazil}

\affil[2]{School of Physics and Astronomy, University of Leeds, Leeds LS2 9JT, United Kingdom}

\affil[3]{International Institute of Physics, Federal University of Rio Grande do Norte, 59070-405 Natal, Brazil}

\affil[4]{Institute for Quantum Studies, Chapman University, One University Drive, Orange, CA, 92866, USA}

\date{}                     
\usepackage{soul}

\begin{document}

\maketitle

\begin{abstract}

Inspired by the one-hundredth anniversary of the seminal works of Stern and Gerlach, our contribution is a proposal of how to use their famous experiment in a more contemporary perspective. Our main idea is to re-cast the experiment in the modern language of prepare-and-measure scenarios. By doing so, it is possible to connect geometric and algebraic aspects of the space of states with the physical space. We also discuss possible simulations of the SG experiment as well as some experimental properties of the experiment revealed at the statistical level. Merging a more modern perspective with a paradigmatic experiment, we hope this paper can serve as an entry door for quantum information theory and the foundations of quantum mechanics.

\end{abstract}

\section{Introduction}

The Stern-Gerlach (SG) experiment branches over all corners of quantum theory. It extends from the very beginning historical aspects of quantum mechanics, concerning the spatial quantization of the angular momentum angle of an atom in relation to a proper magnetic field within the Sommerfeld atomic model \cite{stern_weg_1921} and its follow-up, showing positive results on this the subject \cite{gerlach_experimentelle_1921_1}. The most remembered papers were published a bit later, in 1922 \cite{gerlach_experimentelle_1922_2, gerlach_magnetische_1922}. They are closely attached to the spin discovery, although it was not clear at that time \cite{gomes_o_2011}. Currently, it is the basis for state-of-art measurements on qubits \cite{wu_sterngerlach_2019}.  

Essentially, the experiment consists of sending a beam of particles through an inhomogeneous magnetic field. Simply put, classical descriptions predict some sort of continuous Gaussian-spread profile for the outcomes of this experiment. Contrary to the classical prediction, the beam is not located in one single spot: it is split into two \cite{cohen-tannoudji_quantum_2020}. 

Even though the SG experiment has been  deeply discussed in ref. \cite{sakurai_modern_2017}, in this contribution, however, we would like to use it as a complete tool kit to analyze its very quantum nature, putting aside technical or experimental details. We propose a couple of further investigations. Namely, 
\begin{enumerate}
    \item A concatenation of two SG experiments is a physical proxy for what is called a ``prepare-and-measure'' scenario \cite{gallego_device-independent_2010,GoisEtAl21}.
    
    \item The space of states in an SG setup is two-dimensional; whose generic element is usually called a qubit ($\cong \mathds{C}^2$ over the field $\mathds{C}$ of complex numbers). Other examples, such as photon polarization or even the double slit experiment, are on equal footing. For this case, it is customary to write an arbitrary vector using the standard state parametrization \cite{nielsen.chuang.2011}
    \begin{equation}\label{bloch_sphere_parametrization}
        \ket{\psi} = \cos \frac{\te}{2} \ket{0} + e^{i \varphi} \sin\frac{\te}{2} \ket{1}. 
    \end{equation}
    with $\te \in [0, \pi]$ and $\varphi \in [0, 2 \pi)$. In this case, the vector states lie in what is called a Bloch sphere. Our manuscript provides an operational meaning for the parameters $\te$ and $\varphi$. Rather than a mere mathematical parametrization, the pair $(\te, \varphi)$ identifies the orientation of the magnet used in the preparation of states on the three-dimensional physical space, giving rise to \eqref{bloch_sphere_parametrization}. This intrinsic connection between the three-dimensional (real) space and the two-dimensional (complex) space of states is explored through the Hopf fibration. We discuss it in detail in Appendix A.  
    
    \item Finally, and not less important, we address the current problem of dimension witnesses \cite{PhysRevLett.110.150501, brunner_bell_2014}. The main goal of this research area consists of determining how many degrees of freedom an unknown physical system has only looked at the measured data. For the particular representation of quantum systems with Hilbert spaces, it means to find its dimension given a set of some conditional probabilities or quantum correlations. We will show that, for the canonical case of an SG setup, the dimension $d=2$ is recovered when one tights a particular quantum witness. 
    
\end{enumerate}

In light of the historical and technical comments above, we put this manuscript forward not only to celebrate one hundred years of the seminal SG works but also to explore its pedagogical potential. Recently, the pedagogical approach to quantum mechanics has been found to be conducted through two main lines: one can focus on spin first \cite{corinne}, or one can give more emphasis on wave function first \cite{nathan}. Our work is based on the former. By focusing on spin first, fundamental aspects of quantum theory turn out to be more transparent. For instance, in the SG case, superposition is one of these key aspects that transparently stand out. As we will discuss in more detail below, it is possible to re-frame the SG experiment in a more modern scenario. By emphasizing in a device-independent manner the preparation and the measurements, this re-framing abstracts away the experimental complexities surrounding the original SG device. Focusing exclusively on toy models for preparations and measurements gives the natural framework to design protocols of quantum communication and quantum cryptography — both on the research level and on the pedagogical level. For instance, one can easily understand quantum teleporting \cite{teleporting} or even more paradigmatic algorithms of quantum cryptography (like the famous BB84 \cite{bb84}) under this prepare-and-measure scenario.

The paper is divided as follows. Section \ref{sec.2} is dedicated to introducing the prepare-and-measure scenario, modulated by a concatenation of two SG devices. In Section \ref{sec.3}, we explore some experimental facts related to the scheme previously presented. Leveraging on geometrical grounds, we construct the Bloch sphere in Sec. \ref{sec.4}. We also treat the geometry of probabilities intrinsically connected with quantum mechanics. In Section \ref{sec.5} we address the modern problem of dimension witnesses within the SG experiment. Our main result consists of showing how only the statistical data extracted from a black box scenario leads to the description of the SG as a qubit. Finally, Sec. \ref{sec.6} is left for the conclusions. 

\section{Simulating a prepare-and-measure scenario}
\label{sec.2}

\begin{figure}
    \centering
    \includegraphics[scale=0.24]{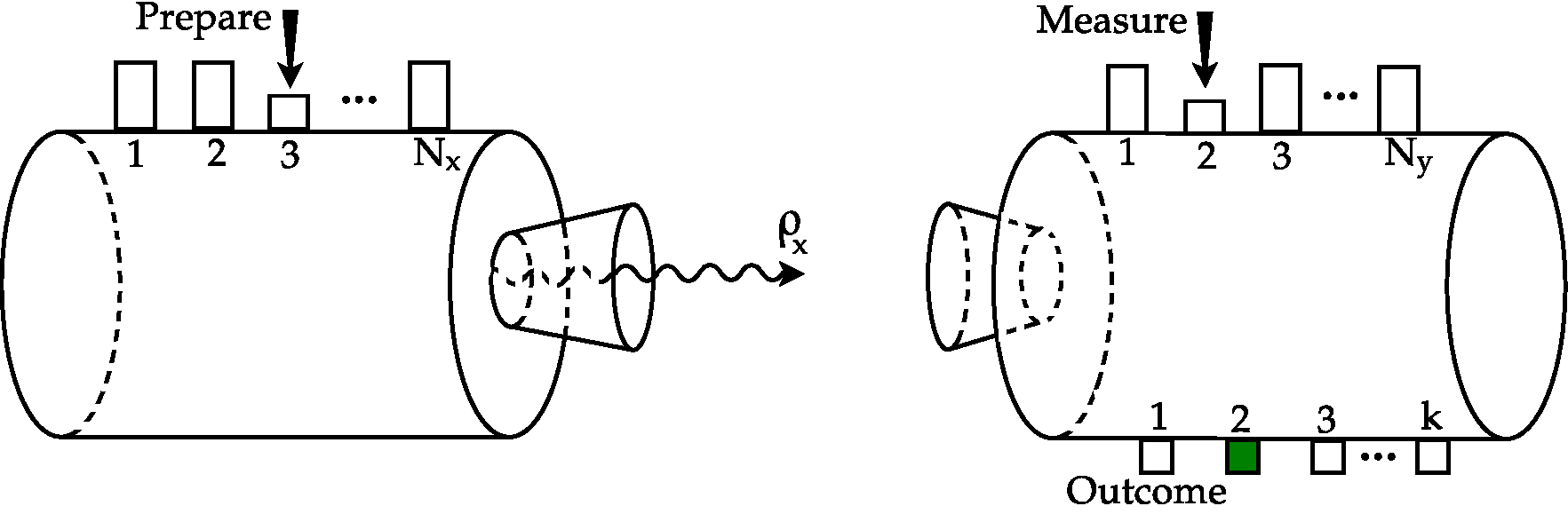}
    \caption{Schematic representation of a prepare-and-measure scenario, with an outcome $a=2$.}
    \label{prep_meas}
\end{figure}

The so-called prepare-and-measure scenario basically consists of two boxes~\cite{gallego_device-independent_2010,GoisEtAl21,DA18}. The first one has $N_x$ buttons that prepare the system in a state $\rho_x$ under demand. The second box has other $N_y$ buttons that perform measurements, providing different outcomes labelled by $a=1,...,k$. Fig. \ref{prep_meas} depicts this idea. 

When no more information about the nature of the physical systems involved in this scenario is available to the experimentalist, the only thing they can do is describe it through a set of conditional probabilities $Pr(a|x,y)$ of getting an outcome $a$, after preparing the system with $x$ and measuring $y$. 

Note that this is exactly what a sequence of two SG devices provides. In fact, to push a button $x$ means, operationally, to select a direction to the magnetic field that separates the flux of particles in two, together with blocking one of the resultant split beams, while the other is free to move on (into the measurement device). Care must be taken when we say ``the direction'' of the magnetic field in this case. One possible realization of the inhomogeneous magnetic field of the SG device is given by
\begin{equation}
    \vec{B}(\vec{r}) = -\xi x \hat{e}_1 + (B_0+ \xi z)\hat{e}_3
\end{equation}
where $\{\hat{e}_i, i=1,2,3 \}$ is the basis of the physical space of displacement vectors $\mathds{D}$ ($\cong \mathds{R}^3$), with coordinates $x$, $y$ and $z$ \cite{ldb2019}. $B_0$ is the component of $\vec{B}$ in the $z$ direction. $\xi$ represents small deviations in the sense that $|\xi z|$ and $|\xi x|$ are much smaller than $|B_0|$. It guarantees that (i) we have an inhomogeneous magnetic field and (ii) $\nabla \cdot \vec{B} = 0$, as expected. In this case, we say that the magnetic field points towards the $z$ direction, by an abuse of notation. We may also realize what the $y$-button represents. It defines another direction of a magnetic field, that, once again, may split the income beam into two. In one of the exits, there is a wall to stop the outcome beam. On the other, though, there is a Geiger-count type detector. This is the spirit of what once was called a ``Yes-No'' experiment or proposition - the core of  older quantum systems descriptions based on the propositional calculus \cite{jauch_foundations_1968}. In modern terminology, we call it a projective measurement (on a qubit) or a test \cite{barbara.2011.book}. It is specified by a set $\mathcal{M} = \{\ket{\psi_-}, \ket{\psi_+} \}$. If we measure the qubit prepared in the state $\rho_x$, then the possible outcomes are $-$ or $+$, interpreted here as being blocked by the wall or counted by the detector. This structure may also be rephrased, noting that $y$ can be interpreted not only as a measurement but also as a preparation of states. Hence, the measurement of the second box can be restated by the (yes-no) question: once the system is prepared in the state $\rho_x$, is the system in the state $\rho_y$? Questions of the form: ``What is the state of the system?'' seems to be difficult to be approached experimentally~\cite{GT09,CramerEtAl10,Hradil97}.  

\section{Experimental facts concerning the SG experiment}
\label{sec.3}

\begin{figure}
    \centering
    \includegraphics[scale=0.4]{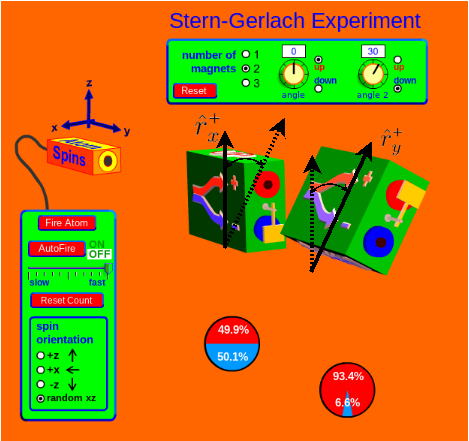}
    \caption{Experimental scheme of two SG devices generated by the PhET interactive simulation.}
    \label{prep_meas_30}
\end{figure}

\begin{figure}
    \centering
    \includegraphics[scale=0.4]{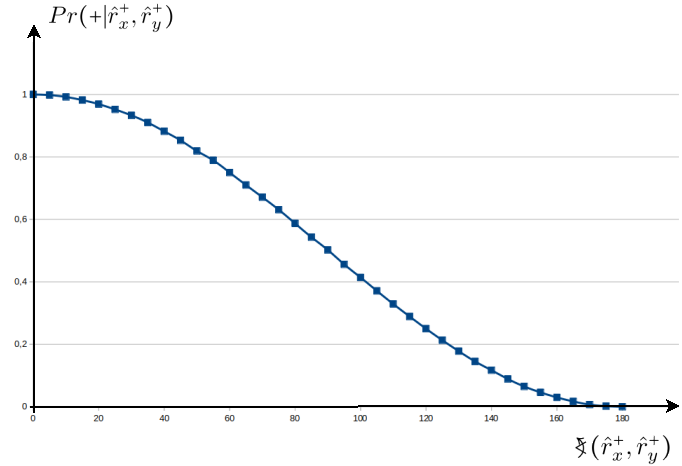}
    \caption{$Pr(+|\hat{r}^+_x, \hat{r}^+_y)$ as a function of the angle between $\hat{r}_x$ and $\hat{r}_y$.}
    \label{graph}
\end{figure}

The basic experimental fact concerning the sequence of two SG devices in the prepare-and-measure scenario is related to a conditional probability. We will use the following notation. Each $x$ (or $y$) button of preparation (or measure) is defined by a direction $\hat{r}_x$ ($\hat{r}_y$) in the physical space, that orients the magnetic field. Moreover, it also selects either one ($+$) or the other ($-$) split beam (both in $x$, allowing one of the scattered beams to head into the measuring device and $y$, to set the detector). So, we define
\begin{equation}
    x \longleftrightarrow \hat{r}^+_x \,\,\mbox{or} \,\, \hat{r}^-_x.
\end{equation}
This notation is explored\footnote{We extensively use the PhET interactive simulations available at \url{https://phet.colorado.edu/pt/simulation/legacy/stern-gerlach}.} in Fig. \ref{prep_meas_30}. In this case, $\hat{r}^+_x$ selects the beam with spin up in the $z$ direction. Else, $\hat{r}^+_y$ counts the atoms that fire the wall with spin up in a direction rotated by $\pi/6$ of the first magnetic field.    
This figure also shows a particular example of an experimental fact,
\begin{equation}\label{pr}
    Pr(+|\hat{r}^+_x, \hat{r}^+_y) = \frac{1+ \cos \arcangleru (\hat{r}^+_x, \hat{r}^+_y)}{2}.
\end{equation}
To see this, we have plotted the graph \ref{graph}, with a repeated sequence of SG experiments, keeping $\hat{r}_x$ fixed while varying $\hat{r}_y$. The curious result is consistent with \eqref{pr}. The data used to plot the graph was, once again, collected using the SG PhET interactive simulations: it is possible to define the angle between the magnetic fields in both devices. 

Before moving on to complete the construction of states in this particular experiment, let us discuss a peculiar feature of a concatenation of SG devices, which is related to the concept of reproducibility of tests. Let us set a sequence of two SG devices, fully characterized by $\hat{r}^+$. We find that $Pr(+|\hat{r}^+, \hat{r}^+)=1$. In other words, the question of having the states prepared and measured by the same magnetic field is answered with $100\%$ of certainty. Furthermore, we could insert another box after the two ones, asking the same question. The answer would be ``yes'' repeatedly. Now comes the tricky step, which has no immediate classical counterpart~\cite{LJ15}. Let us consider three SG devices, characterized by, say, $\hat{r}^+$, $\hat{u}^+$ and $\hat{r}^+$, respectively. Each box may be interpreted as either prepare or measuring device. This is so because the income beam, as previously discussed, is divided by two. The first and the third are the same. However, the second test destroys the preparation performed by the first box. Thus, the answer in the third one is not responded to repeatedly. In this sense, the first and second tests are called incompatible. Conversely, two distinct tests, say $\mathcal{A}$ and $\mathcal{B}$ are called compatible whenever the test $\mathcal{B}$, applied in between two repetitions of $\mathcal{A}$, does not affect the reproducibility of $\mathcal{A}$. We point out that this situation is drastically different from its classical counterpart. In fact, for a spinning top, with angular momentum $\vec{L} = I \vec{\omega}$, one could obtain, say $\omega_x$ and $\omega_y$ simultaneously. Here, $I$ stands for the moment of inertia, and $\vec{\omega}$ is the corresponding angular velocity \cite{sakurai_modern_2017}.

\section{Bloch sphere unveiled}
\label{sec.4}

The discussion presented so far hasn't revealed the proper representation of states in the SG experiment. The incompatibility of tests discussed before is just a piece of evidence for its quantum description. Besides that, the division of the income beam in two indicates that we could use a two-dimensional Hilbert space to fully represent the experiment. Let us assume, then, that the state of space is just $\mathds{C}^2$, with vectors denoted by the standard Dirac notation. This \textit{ad hoc} imposition leads to the correct conditional probability \eqref{pr} through a Born rule, to be presented in a while, see \eqref{born1} and \eqref{born2}. We will also construct the Bloch sphere here. To do so, we divide our tasks into subsections. In the first one, we try to explore a liaison between the geometry of Hilbert spaces and probabilities. Then, we direct our efforts to the Bloch sphere itself. Finally, a list of comments will be given.  

\subsection{Geometry of probabilities}

The parametrization given in \eqref{bloch_sphere_parametrization} reflects a clean geometrical meaning in terms of the preparation of states. In fact, the pair $(\theta, \varphi)$ uniquely describes the state prepared by 
\begin{equation}\label{xrplus}
    x \longleftrightarrow \hat{r}^+_x=\sin \theta \cos \varphi \hat{e}_1 + \sin \theta \sin \varphi \hat{e}_2 + \cos \theta \hat{e}_3, 
\end{equation}
that is, the radial arbitrary direction in spherical coordinates in the physical space that indicates the magnetic field orientation. As usual, $\{\hat{e}_i, i=1,2,3\}$ is the canonical basis of $\mathds{R}^3$.  

Let us clarify this point. As mentioned before, the two split beams indicate that $\mathds{C}^2$ is a good candidate for the space of states. Let us choose a basis in $\mathds{C}^2$. By pressing the $x$-button of preparation, we select a particular direction in space to point the magnetic field as well as one of the scattered beams. So, we take
\begin{eqnarray}\label{x1x2}
\begin{matrix}
x_1 \longleftrightarrow \hat{e}^+_3 \longleftrightarrow \ket{0}, \\
x_2 \longleftrightarrow \hat{e}^-_3 \longleftrightarrow \ket{1}.
\end{matrix}
\end{eqnarray}
It means that the vectors $\ket{0}$ and $\ket{1}$ represent the two possible preparations. These two possible choices mean that the system may bear excluding properties (\textit{i.e.}, of being selected by $x_1$ or $x_2$ in \eqref{x1x2}). So, we associate a property of the system to a subspace in the state of space. There is one way to describe what is the probability of a system to posses a particular property: we look at the projection of the state vector onto the corresponding subspace. This geometry of probabilities asks for a inner-product in $\mathds{C}^2$, $\braket{ \vert }: \mathds{C}^2 \times \mathds{C}^2 \rightarrow \mathds{C}$. For our case, if the system is prepared in the state $\ket{1}$, it should have no component in the subspace spanned by $\ket{0}$. In fact, the properties of being selected by $x_1$ or $x_2$ are excluded. Thus, $\bra{0} \ket{1}=0$, which also implies that $\mathcal{Z} = \{ \ket{0}, \ket{1} \}$ can be taken as a basis for $\mathds{C}^2$, as they are linearly independent. Along with the lines above, for a general state vector
\begin{equation}\label{alphabeta}
    \ket{\psi} = \alpha \ket{0} + \beta \ket{1},
\end{equation}
where $|\alpha|^2$ represents the probability of the system to be measured by $y_1 \longleftrightarrow \hat{e}^+_3$ as well as $|\beta|^2$ by $y_2 \longleftrightarrow \hat{e}^-_3$, owing to the geometry of probabilities so constructed. 

Mathematically, we project $\ket{\psi}$ on the corresponding subspace and take the square. That is, if $\hat{r}^+ \longleftrightarrow \ket{\psi} = \alpha \ket{0} + \beta \ket{1}$, then
\begin{equation}\label{probab+-}
\begin{matrix}
    Pr(+|\hat{r}^+, \hat{e}^+_3) = |\bra{0} \ket{\psi}|^2 = |\alpha|^2, \\
    Pr(-|\hat{r}^+, \hat{e}^-_3) = |\bra{1} \ket{\psi}|^2 = |\beta|^2.
\end{matrix}    
\end{equation}
Instead of $\ket{0}$ and $\ket{1}$, one could take projectors to represent states, 
\begin{equation}
    \rho_{x_1} = \ket{0} \bra{0}, \, \, \, \rho_{x_2} = \ket{1} \bra{1}, 
\end{equation}
or even $\rho_x = \ket{\psi} \bra{\psi}$. The probabilities estimated in \eqref{probab+-} assume the form
\begin{equation}\label{Tr}
\begin{matrix}
Pr(+|\hat{r}^+, \hat{e}^+_3) = Tr(\rho_x \rho_{x_1}) = |\alpha|^2, \\
Pr(-|\hat{r}^+, \hat{e}^-_3) = Tr(\rho_x \rho_{x_2}) = |\beta|^2.
\end{matrix}
\end{equation}
In this case, one or the other option will be answered. So, $|\alpha|^2+|\beta|^2=1$ has a clear meaning in terms of probabilities. Another possible geometrical origin to quantum probabilities (and their difference to the classical ones) may be found in \cite{isham}.

\subsection{Geometrical interpretation to the Bloch sphere}

Our final task consists of finding the complex coefficients $\alpha$
and $\beta$ in terms of the duple $(\theta, \varphi)$ that defines $\hat{r}^+$ uniquely, see \eqref{xrplus}. 

Firstly, we may guess what are the states corresponding to the preparations $\hat{e}^{\pm}_1$. According to $Pr(+|\hat{e}^+_1, \hat{e}^+_3)= Pr(-|\hat{e}^-_1, \hat{e}^-_3) = \frac{1}{2}$, the prepared states oscillate when asked about $\hat{e}^+_3$. Thus,
\begin{equation}
    \begin{matrix}
    \hat{e}^{+}_1 \longleftrightarrow \frac{1}{\sqrt{2}}(\ket{0} + \ket{1})=: \ket{+}, \\
    \hat{e}^{-}_1 \longleftrightarrow \frac{1}{\sqrt{2}}(\ket{0} - \ket{1})=: \ket{-},
    \end{matrix}
\end{equation}
and $\braket{+\vert -} = 0$, as expected (the properties of being prepared by $\hat{e}^+_1$ and $\hat{e}^-_1$ are excluding). The set $\mathcal{X} = \{\ket{+}, \ket{-} \}$ could also be taken as a basis for $\mathds{C}^2$. The change $\mathcal{Z} \to  \mathcal{X}$ can be seen as a rotation of $\pi/4$ (or a unitary transformation in the case of complex vector spaces). Likewise, we can ``rotate'' in the complex state of spaces both $\ket{0}$ and $\ket{1}$ by a $\pi/4$ factor to generate the basis $\mathcal{Y} = \{ \ket{a}, \ket{b} \}$. First, we write 
\begin{equation}
\begin{matrix}
    \ket{a'} = \frac{1}{\sqrt{2}}(e^{i \frac{\pi}{4}}\ket{0} + e^{-i \frac{\pi}{4}}\ket{1}), \\
    \ket{b'} = \frac{1}{\sqrt{2}}(e^{i \frac{\pi}{4}}\ket{0} - e^{-i \frac{\pi}{4}}\ket{1}).
\end{matrix}    
\end{equation}
Noting that a global phase factor is irrelevant (in the sense of preserving probabilities), we finally define
\begin{equation}
\begin{matrix}
    \ket{a} = e^{-i \frac{\pi}{4}}\ket{a'} = \frac{1}{\sqrt{2}}(\ket{0} - i \ket{1}), \\ \ket{b} = e^{-i \frac{\pi}{4}}\ket{b'} = \frac{1}{\sqrt{2}}(\ket{0} + i \ket{1}).
    \end{matrix}
\end{equation}
As in the case of $\mathcal{Z}$ and $\mathcal{X}$, $\mathcal{Y}$ is also formed by orthonormal vectors and 
\begin{equation}
    \hat{e}^{\pm}_2 \longleftrightarrow \frac{1}{\sqrt{2}}(\ket{0} \pm i \ket{1}).
\end{equation}
They generate the correct oscillations of $50\%/50\%$ in the probabilities,
\begin{equation}
    Pr(\pm| \hat{e}^+_i, \hat{e}^{\pm}_j) = \frac{1}{2}, \, \, i,j=1,2,3, \, i \neq j.
\end{equation}

With the basis $\mathcal{X}$, $\mathcal{Y}$ and $\mathcal{Z}$ we are in position to obtain the coefficients $\alpha$ and $\beta$ in \eqref{alphabeta}, such that 
$\hat{r}^+ \longleftrightarrow \ket{\psi}$ and $\hat{r}$ is an arbitrary direction in the physical space. 

Invoking once again that the conditional probability \eqref{pr} is only a function of the angles between the magnetic field vectors involved in preparing and measuring, we have
\begin{equation}
    Pr(+| \hat{r}^+, \hat{e}^+_3) = \frac{1+ \hat{r} \cdot \hat{e}_3}{2} = \frac{1+ \cos \theta}{2} = |\alpha|^2.
\end{equation}
The last equality holds due to the geometry of probabilities constructed so far. So, $\alpha = \cos \frac{\theta}{2} e^{i\lambda}, \, \lambda \in \mathds{R}$. Since a global phase factor does not influence probabilities, we hide $e^{i\lambda}$ in $\beta$. 

Following this spirit and knowing that 
\begin{equation}
\begin{matrix}
\hat{e}^+_1 \longleftrightarrow \ket{+} = \frac{1}{\sqrt{2}}(\ket{0} + \ket{1}), \\
\hat{e}^+_2 \longleftrightarrow \ket{a}= \frac{1}{\sqrt{2}}(\ket{0} +i \ket{1}),
\end{matrix}
\end{equation}
we have
\begin{equation}
\begin{matrix}
    Pr(+|\hat{r}^+, \hat{e}^+_1) = \frac{1+ \sin{\theta} \cos{\varphi}}{2} = |\bra{+}\ket{\psi}|^2  \Rightarrow  \\  \Rightarrow 
    1+\sin{\theta} \cos{\varphi} = \cos^2{\frac{\theta}{2}}+ \cos{\frac{\theta}{2}}(\beta + \beta^*) + |\beta|^2.
\end{matrix}
\end{equation}
Since $|\beta|^2 = 1 - |\alpha|^2 = 1 - \cos^2{\frac{\theta}{2}}$, we find, 
\begin{equation}
    Re(\beta) = \sin{\frac{\theta}{2}} \cos{\varphi}.
\end{equation}
Analogously, 
\begin{equation}
    Pr(+|\hat{r}^+, \hat{e}^+_2) = \frac{1+ \sin{\theta} \sin{\varphi}}{2} = |\braket{a \vert \psi}|^2
\end{equation}
implies 
\begin{equation}
    Im(\beta) = \sin{\frac{\theta}{2}} \sin{\varphi}.
\end{equation}
If we now gather all the pieces, then $b = \sin{\frac{\theta}{2}} e^{i \varphi}$ and 
\begin{equation}\label{b}
    \hat{r}^+ \longleftrightarrow \ket{\psi} = \cos\frac{\theta}{2} \ket{0} + e^{i \varphi} \sin \frac{\theta}{2} \ket{1},
\end{equation}
completing our final objective of operationally constructing \eqref{bloch_sphere_parametrization}. To avoid a lengthy discussion here, we leave the next subsection for general comments and discussions concerning our approach.     

\subsection{General comments and discussions}

\begin{enumerate}
    \item Starting from a three-dimensional physical space, we may construct a two-dimensional state of spaces. Actually, we assumed that we had a two-dimensional Hilbert space. This assumption has led us to a theoretical construction that fits perfectly with the experimental data available. What can be said about the inverse? The existence of qubits necessarily needs a three-dimensional space?
    
    \item Our construction has shown a geometrical significance to the pair $(\theta, \varphi)$ used to parametrize state vectors, given by \eqref{b}. If the pair runs values in the standard domain $\theta \in [0, \pi]$ and $\varphi \in [0, 2 \pi]$, then we cover the unit sphere, and so does $\ket{\psi}$ over what is called the Bloch sphere.  
    With a fixed $\varphi=0$, the two vectors with $\theta = 0$ and $\theta = \pi$ are connected with the following states, 
    \begin{equation}
        \begin{matrix}
        \hat{r}^+_1 = \hat{e}^+_3 \longleftrightarrow \ket{0}, \\
        \hat{r}^+_2 = - \hat{e}^+_3 \longleftrightarrow \ket{1}. 
        \end{matrix}
    \end{equation}
This calculation indicates a general result: antipodes on the unit sphere $S^2$ are related to orthonormal states. In fact, in physical space, the pairs $(\theta, \varphi)$ and $(\pi - \theta, \varphi+ \pi)$ generate a couple of antipodes and, as stated, we have
\begin{equation}
    \begin{matrix}
    \hat{r}^+(\theta, \varphi) \longleftrightarrow \ket{\psi_+} = \cos{\frac{\theta}{2}} \ket{0} + e^{i \varphi} \sin{\frac{\theta}{2}} \ket{1}, \\
    \hat{r}^-(\theta, \varphi) \longleftrightarrow \ket{\psi_-} = \sin{\frac{\theta}{2}} \ket{0} - e^{i \varphi} \cos{\frac{\theta}{2}} \ket{1}.
    \end{matrix}
\end{equation}
where
\begin{equation}
    \hat{r}^+(\theta, \varphi) = \sin \theta \cos \varphi \hat{e}_1 + \sin \theta \sin \varphi \hat{e}_2 + \cos \theta \hat{e}_3, 
\end{equation}
and $\hat{r}^-(\theta, \varphi) =  \hat{r}^+(\pi- \theta, \varphi+ \pi)$. A direct calculation shows that $\hat{r}^+ \cdot \hat{r}^- = -1$ as well as $\braket{\psi_+ \vert \psi_-}=0$. Figure \ref{s2} shows a geometrical picture of this result in both the unit sphere $S^2$ immersed in the physical space (denoted by $\mathds{D}$) and the Bloch one. 
\begin{figure}
    \centering
    \includegraphics[scale=0.34]{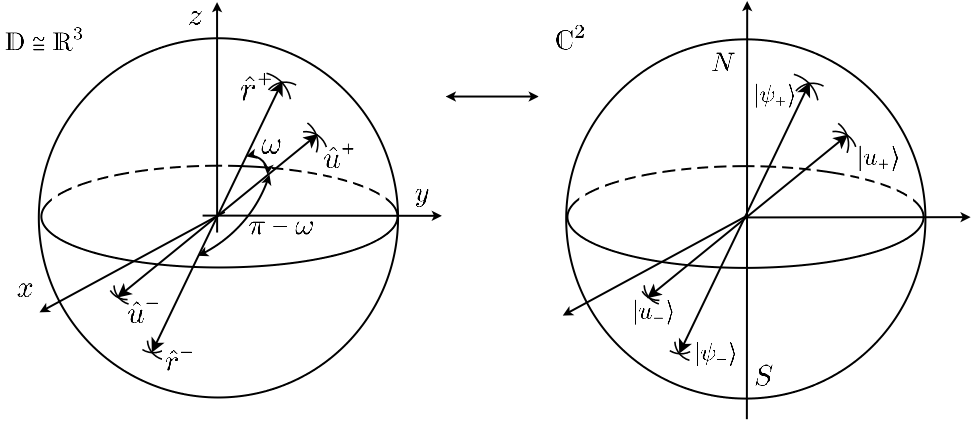}
    \caption{Geometrical representation of $S^2$ and the Bloch sphere: antipodes represent orthonormal states.}
    \label{s2}
\end{figure}

We may also explore Figure \ref{s2} to give some perspective on the geometry of probabilities that we spoke about before. Suppose that $\hat{u}$ is a unitary vector such that the system is prepared in the state
\begin{equation}
    \hat{u}^+ \longleftrightarrow \ket{u_+}.
\end{equation}
We shall perform a test designated by $\mathcal{M} = \{\ket{\psi_+}, \ket{\psi_-} \}$. When $\ket{u_+}$ falls in the north hemisphere (denoted by an $N$ in Fig. \ref{s2}), it is more likely to find a $+$ in the measurement. If $\ket{u_+}$ is such that the angle between $\hat{r}$ and $\vec{u}$ is $\pi/2$, then the probabilities of getting $+$ or $-$ are the same and equal $1/2$. To see it, we set $\arcangleru (\hat{u}, \hat{r}) = \omega$. Thus, 
\begin{eqnarray}\label{p1}
Pr(+|\hat{u}^+, \hat{r}^+) &=& \frac{1}{2}(1 + \cos \omega) = \cos^2 \left ( \frac{\omega}{2} \right ) \\ \label{p2}
    Pr(-|\hat{u}^+, \hat{r}^-) &=& \frac{1}{2}(1 + \cos{(\pi- \omega})) = \sin^2 \left ( \frac{\omega}{2} \right )
\end{eqnarray}

Although we depict $\ket{\psi_+}$ and $\ket{\psi_-}$ falling in the same straight line, in opposite directions, they span orthogonal subspaces in $\mathds{C}^2$, $span \{\ket{\psi_+} \}^{\perp} = span \{\ket{\psi_-} \}$. With our limited vision, the probability values in \eqref{p1} and \eqref{p2} would suggest sketching $\ket{\psi_+}$ and $\ket{\psi_-}$ with an angle of $\pi/2$, contrary to what was shown in the Bloch sphere of Fig. \ref{s2}. Unfortunately, it is the best we can do.    

\item Previously, we have introduced the use of traces to evaluate probabilities; see \eqref{Tr}. Let us expand this idea, which shall align our notation to the next Section. We now invert what we have done in the previous item: suppose that the system is prepared in the state $\rho_x = \ket{\psi_+}\bra{\psi_+}$, where $\ket{\psi_+}$ was defined in \eqref{b}. A direct calculation shows that
\begin{equation}
    \rho_x = \frac{1}{2}( 1\!\!1_{2}+ \hat{r} \cdot \vec{\sigma}).
\end{equation}
Here, $1\!\!1_{2}$ is the identity matrix of order $2$.  The calculation was carried out in the canonical representation 
\begin{equation}
    \ket{0}= \begin{pmatrix} 1 \\ 0
    \end{pmatrix}, \,\,
    \ket{1}= \begin{pmatrix} 0 \\ 1
    \end{pmatrix}.
\end{equation}
$\vec{\sigma} = (\sigma_x, \sigma_y, \sigma_z)$ are the Pauli matrices and, in this case, assume the form,
\begin{equation}
    \sigma_x = \begin{pmatrix}
    0 & 1 \\
    1 & 0
    \end{pmatrix}, \quad
    \sigma_y = \begin{pmatrix}
    0 & -i \\
    i & 0
    \end{pmatrix}, \quad
    \sigma_z = \begin{pmatrix}
    1 & 0 \\
    0 & -1
    \end{pmatrix}.
\end{equation}

The measurement, in turn, is associated to $\mathcal{M} = \{\ket{u_+}, \ket{u_-} \}$ where $\hat{u}^+$ is obtained by a shift of $\omega$ in the $\theta$-coordinate of $\hat{r}^+$. In this case, 
\begin{eqnarray}
\ket{u^+} = \cos\left (\frac{\theta+ \omega}{2} \right )\ket{0} + e^{i\varphi}\sin\left (\frac{\theta+ \omega}{2} \right )\ket{1}, \\
\ket{u^-} = \sin \left (\frac{\theta+ \omega}{2} \right )\ket{0} - e^{i\varphi}\cos \left (\frac{\theta+ \omega}{2} \right )\ket{1}.
\end{eqnarray}
A direct calculation shows the following results,
\begin{eqnarray}\label{born1}
    Pr(+| \hat{r}^+, \hat{u}^{+}) = Tr(\rho_x \mathcal{M}_{+})= \cos^2 \left ( \frac{\omega}{2} \right ), \\ \label{born2}
    Pr(-| \hat{r}^+, \hat{u}^{-}) = Tr(\rho_x \mathcal{M}_{-})= \sin^2 \left ( \frac{\omega}{2} \right ),
\end{eqnarray}
where $\mathcal{M}_{\pm} = \ket{u^{\pm}}\bra{u^{\pm}}$ are orthogonal projectors. They play a role in this particular case of a broader class of operators that can be used to generalize what was defined in \eqref{born1} and \eqref{born2}. In general, it is said that a system has a quantum behaviour when probabilities can be expressed by a Born rule,
\begin{equation}
Pr(a|x,y) = Tr(\rho_x \mathcal{M}^a_y),
\end{equation}
where $a$ means the result of a test, $x$ and $y$ stands for preparation and measurement and $\mathcal{M}^a_y$ is a positive operator-valued measure (POVM). Our entire analysis shows that the SG falls in this type of description, as the standard example of a quantum system, being described accordingly.
\end{enumerate}

\section{Witnessing Dimensions in the Stern-Gerlach}
\label{sec.5}

Until this point, we have tacitly assumed that the space of states used to make sense of the physics involved in the Stern-Gerlach experiment is two-dimensional. Incidentally, the choice of a qubit can be seen as nothing but a mere artefact. Two quantum degrees of freedom are presumably enough to explain the aggregated statistics arising out of several rounds of preparations and measurements in the Stern-Gerlach scenario. Put another way, the agreement with the probabilistic predictions could justify adopting a two-dimensional Hilbert space as it is the simplest explanation for the experiment.

Nonetheless, the previous section hints that there may be a deeper link connecting our three-dimensional physical space with the space of states for qubits. The question is, can we invert this situation? Put another way, can we use only experimental data to infer the underlying space of states? This is the central problem of what is known as dimension witnesses~\cite{PhysRevLett.110.150501,CBRM16,KV17,Brunner2008,gallego_device-independent_2010}.   

Imagine the situation where we perform two SG in sequence, but in a way that no information about the experiment is available anywhere other than for the third party that has prepared the setup. Much in the spirit of refs.~\cite{gallego_device-independent_2010,Brunner2008}, in this situation, the magnets can be cast as truly black-boxes that prepare and measure on demand the system under investigation. More precisely, we could say that the system is prepared in the state $\rho_x$ by pressing the button $x \in \{1,2,...,N \}$. After being prepared, the system is measured by selecting a certain $y\in \{1,...,m \}$. Here, $N$ and $m$ enumerate the number of buttons available on each box - see fig.~\ref{prep_meas}. After a certain measurement is selected, one particular outcome is recorded: named $a$. The aggregated statistics coming out of this experimental setup is naturally recovered via the Born rule 
\begin{equation}
Pr(a|x,y) = Tr(\rho_x \mathcal{M}^a_y),
    \label{Eq.BornRule}
\end{equation}
where $\{\mathcal{M}_{y}^{a}\}_{a}$ is a POVM for each choice of measurement button. Finally, we say that $f(\cdot)$ is a \emph{quantum dimension witness} when it is upper-bounded
\begin{equation}
    f\left (Pr(a|x,y) \right ) \leq Q_d 
\end{equation}
for all experiments involving quantum systems of Hilbert space dimension no greater $d$~\cite{PhysRevLett.110.150501}. Before moving on, let us consider an example. On the scenario so described in Fig. \ref{prep_meas}, consider the particular and simple case in which there are $N$ possible preparations and just one measurement, that is, $m=1$. We may construct a dimension witness with the average probability
\begin{equation}
    U_N = \frac{1}{N}\sum^N_{x=1} Pr\left (b=x \vert x \right ).
\end{equation}
We can promptly obtain a superior bound on $U_N$, as a function of $d$. In fact, due to the properties of the density operator $\rho_x$, 
\begin{equation}\label{testemunho}
    U_N = \frac{1}{N}\sum^N_{x=1} Tr\left ( \rho_x \mathcal{M}_x \right ) \leq \frac{1}{N} \sum_x Tr\left ( \mathcal{M}_x \right ) = \frac{d}{N} = Q_d. 
\end{equation}
Thus, we are led to a dimension witness for any $d<N$.

In our imagined Stern-Gerlach in a black-box scenario, there are $m=\frac{N(N-1)}{2}$ dichotomic measurements with possible outcomes labelled $\pm 1$. A possible dimension witness can be constructed as follows~\cite{PhysRevLett.110.150501}: 
\begin{equation}
    W_N= \sum_{x>x'}\vert Pr(x,y)- Pr(x',y) \vert^2.
    \label{Eq.OurWitnessNandP}
\end{equation}
We are using the notation $y=(x,x')$ for each measurement and $Pr(x,(x,x')):=Pr(b=1 \vert x, y)$. Accordingly, $\mathcal{M}_{(x,x')}$ is the associated POVM corresponding to the outcome $b=1$. Due to the inequality in \eqref{testemunho}, and given that $W_N$ is a difference of probabilities, one may also write
\begin{equation}\label{wnmenorquedovern}
    W_N \leq Q_d = \frac{d}{N}.
\end{equation}
This restriction will be useful in a while. 

We can rewrite eq.~\eqref{Eq.OurWitnessNandP} as
\begin{equation}
    W_N = \sum_{x>x'} \vert \mbox{Tr}[(\rho_x - \rho_{x'})\mathcal{M}_{(x,x')}]  \vert^2 \leq \sum_{x>x'}\vert D(\rho_x, \rho_{x'}) \vert^2,
\end{equation}
where $D(\rho_x, \rho_{x'})$ stands for the \emph{trace distance} between two density operators:
\begin{equation}
    D(\rho_x, \rho_{x'}):= \max_{\mathcal{M}}\{ \mbox{Tr}[(\rho_x - \rho_{x'})\mathcal{M} \},
\end{equation}
and operationally represents how well two quantum states can be distinguished from each other when allowing for the most general measurement. Alternatively, the trace distance may also be related to another function of distinguishability, namely, the \emph{fidelity} defined by 
\begin{equation}
    F(\rho_x, \rho_{x'}):= \mbox{Tr}\left ( \sqrt{ \sqrt{\rho_{x}} \rho_{x'} \sqrt{\rho_{x}}} \right ).
\end{equation}
Due to the Fuch-van de Graaf inequalities \cite{watrous_theory_2018}, we have 
\begin{equation}
    1-D(\rho_x, \rho_{x'}) \leq F(\rho_x, \rho_{x'}) \leq \sqrt{1-D^2(\rho_x, \rho_{x'})}.
\end{equation}
The second inequality is the one that interests us the most. It allows writing
\begin{equation}\label{43}
    W_N \leq \sum_{x>x'}(1-F^2(\rho_x, \rho_{x'}))=\sum_{x>x'}(1-\vert \braket{\Psi_x \vert \Psi_x'}\vert^2)
\end{equation}
for pure states $\rho_x = \ket{\Psi_x} \bra{\Psi_x}$. Now we write
\begin{align}\label{44.abcd}
    \sum_{x>x'}\vert \braket{\Psi_x \vert \Psi_x'} \vert^2 &=\frac{1}{2}\left ( \sum_{x,x'}\vert \braket{\Psi_x \vert \Psi_x'} \vert^2 - N \right )  \nonumber \\
    &=\frac{N^2}{2}Tr(\Omega^2)- \frac{N}{2},
\end{align}
where we define $\Omega:= \frac{1}{N} \sum^N_{x=1}\ket{\Psi_x} \bra{\Psi_x}$.

We can now insert the dimension of the attached Hilbert space by noting that 
\begin{equation}
    Tr(\Omega^2)\geq \frac{1}{d}
\end{equation}
holds for any normalized state $\Omega$. Finally, the eq. \eqref{43} and eq. \eqref{44.abcd} can be used to derive the inequality below:
\begin{equation}\label{dimension.wit}
    W_N \leq \frac{N^2}{2} \left ( 1- \frac{1}{\min\{ d, N \} } \right ).
\end{equation}

The compelling feature of the above witness is its tightness. It can be shown that for a suitable choice of states $\{\rho_x \}^x_{x=1}$ as well as measurement operators $\mathcal{M}_{(x,x')}$ the ineq. \eqref{dimension.wit} is saturated. We leverage this particular feature in our case. 

Working with the tightness case, the ineq. \eqref{dimension.wit} becomes the following equality, 
\begin{equation}
\frac{d}{N}=\frac{N^2}{2} \left ( 1- \frac{1}{\min\{ d, N \} } \right )
\label{Eq.CentralEquationWitnessDimension}
\end{equation}
where we have used the ineq. \eqref{wnmenorquedovern} and \eqref{dimension.wit}. This can be seen as an equation for obtaining $d$ given that $\mathcal{M}_{(x,x')}$ is the measurement that optimally discriminates between $\ket{\Psi_x}$ and $\ket{\Psi_{x'}}$, whenever $d\leq N$. Due to the simplicity of our experiment, we take $N=2$.  This substitution back in the quadratic equation in the unknown $d$ \eqref{Eq.CentralEquationWitnessDimension} provides the unique solution $d=2$. This result indicates that the underlying quantum system prepared and measured in the Stern-Gerlach experiment must be two-dimensional. Our results could be stretched a bit further, it is not only the case that we are dealing with qubits in the SG experiment, in our regime, they cannot be anything else. 

For arbitrary $N$, if we could open up the black-boxes lid, and look at the inner mechanisms dictating the functioning of the boxes, we would get:
\begin{equation}
W_N = \sum^{N}_{i=1} \left \vert \frac{1 + \cos \theta_i}{2} - \frac{1 + \cos \theta_{i+1}}{2} \right \vert^2 =\frac{1}{4} \sum^{N}_{i=1} \left (  \cos \theta_i - \cos \theta_{i+1}\right )^2      
\end{equation}
by a direct application of the result expressed in eq. \eqref{pr}. As expected, the restriction \eqref{dimension.wit} is obeyed once (i) the difference of cosines is restricted by $1$ and (ii) we are setting $d=2$ in the aforementioned inequality. 

We can also interpret $N$ preparation buttons as selecting an arbitrary direction to the magnetic field in the SG device, and also picking the ``spin up'' preparing the state $\rho_x$. The second box is just another SG device as described previously, see, for example, fig. \ref{prep_meas_30}.

\section{Conclusion}
\label{sec.6}

An entire century has passed since the seminal works of Stern and Gerlach. The motivation for this work was not only to celebrate its one-hundredth anniversary but also to explore the pedagogical potential a two-level system may provide. This way, we have used the SG device to discuss and detail many key topics in modern quantum mechanics. Let us enumerate our main results. 

\begin{enumerate} 
    \item A concatenation of two SG apparatuses may be seen as a proxy to the so-called prepare-and-measure scenario. The first device splits the incoming beam of particles in two and prevents one of the divided beams to move on. In this case, we say that the allowed beam was prepared in the state, say, $\rho$. The second device, in turn, measures the system previously prepared. The usual question it proposes to answer is: ``Is the system in the state $\rho'$?''. The preparation and measurement steps assume a rather concrete form once it is obtained through the operational procedure of allowing the beam of particles to cross a spatial region fulfilled with a magnetic field. 
    
    \item One of the central characteristics of a quantum system is its irreducible probabilistic structure \cite{jauch_foundations_1968}. In our construction, we may literally see it, with collected data from a PhET interactive simulation, see Figs. \ref{prep_meas_30} and \ref{graph}. Actually, the probabilities involved in our formalism can be written in terms of the angle between the magnetic fields involved in preparing and measuring the system, see \eqref{pr}.
    
    \item The geometrical representation of pure states in two-level quantum systems is depicted in the Bloch sphere. The underlying Hilbert space is spanned by the basis $\{\ket{0}, \ket{1} \}$ and an arbitrary vector is written as the linear combination 
    \begin{equation}\label{parametrization.once.again}
        \ket{\psi} = \cos \frac{\te}{2} \ket{0} + e^{i \varphi} \sin\frac{\te}{2} \ket{1}. 
    \end{equation}
    Starting of from a geometry of probabilities, our construction provides a clear geometrical meaning to the parameters $\theta$ and $\varphi$ in \eqref{parametrization.once.again}. They are in one-to-one correspondence with the spherical coordinates in the physical space. When we prepare a system selecting the beam with spin up after passing it through a magnetic field with direction 
    \begin{equation}
        \hat{r}(\theta, \varphi) = \sin \theta \cos \varphi \hat{e}_1 + \sin \theta \sin \varphi \hat{e}_2 + \cos \theta \hat{e}_3,
    \end{equation}
    then the state is represented by $\ket{\psi}$ in \eqref{parametrization.once.again}. This calculation elucidates the deep connection between the three-dimensional physical space and a qubit represented mathematically by a two-dimensional complex space. 
    
    \item Finally, our last result concerns what is called a dimension witness. Basically, given an unknown quantum system, such formalism tries to derive bounds on the dimension of the underlying Hilbert space in order to reproduce the collected measurement data. We have shown that, with a particular black box scenario, two possible preparations are enough to reconstruct a two-dimensional space, consistent with the description of the SG as a true qubit. This conclusion confirms the profound relationship between spatial degrees of freedom and quantum mechanics. There are many more experiments and physical quantities that could be used to explore the intricate relationship between the physical space and the corresponding Hilbert space description of quantum theory, for instance, the polarization of photons. Given the celebratory occasion of the SG setup, in this paper, we are centring our attention exclusively on that physical experiment; we shall address this connection (physical space with Hilbert space) elsewhere, though. At last, in \cite{PhysRevResearch.2.013112}, the authors typify quantum correlations as a function of local symmetries. As stated, it indicates a foundational connection between quantum theory and space-time itself.
\end{enumerate}

\section*{Acknowledgements}

CD thanks the hospitality of the Institute for Quantum Studies at Chapman University. The authors thank immensely the anonymous referee for the valuable suggestions. This work has been supported by Programa Institucional de Bolsas de Iniciação Científica - XXXII BIC/Universidade Federal de Juiz de Fora - 2019/2020, project number 46729 (Programa Institucional de Bolsas de Iniciação Científica - XXXIV BIC/Universidade Federal de Juiz de Fora - 2021/2022, project number 48982).  This research was partially supported by the National Research, Development and Innovation Office of Hungary (NKFIH) through the Quantum Information National Laboratory of Hungary and through the grant FK 135220. This research was also supported by the Fetzer Franklin Fund of the John E.\ Fetzer Memorial Trust and by grant number FQXi-RFP-IPW-1905 from the Foundational Questions Institute and Fetzer Franklin Fund, a donor advised fund of Silicon Valley Community Foundation.  

\section*{Appendix A - The geometry of indistinguishable states}

This Appendix is devoted to formalize the connection between qubits and what is known technically as the Hopf fibration. A pure geometrical/topological exposition may be found in \cite{andre.rafa.impa.2021}. 

\begin{figure}
    \centering
    \includegraphics[scale=0.4]{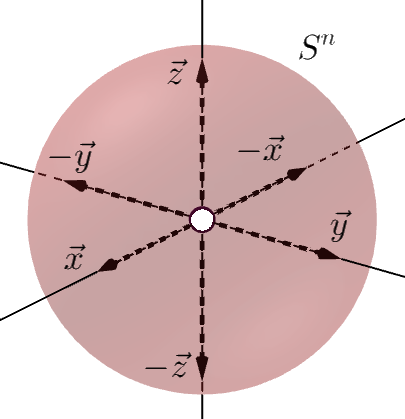}
    \caption{Geometric representation of the unity sphere $S^2$, with identified antipodes.}
    \label{appendix01}
\end{figure}

Our starting point will be the real projective spaces. They can be promptly generalized to complex spaces and, at the same time, admits a clear geometrical meaning. Firstly, we consider the linear space $\mathds{R}^{n+1}$, with elements denoted by $\vec{x}$, $\vec{y}$, etc and $n \in \mathds{N}$. Let $\sim$ be the following relation
\begin{equation}\label{1.1}
    \vec{x} \sim \vec{y} \Leftrightarrow \vec{y} = \lambda \vec{x}; \, 0 \neq \lambda \in \mathds{R}.
\end{equation}
A direct verification shows that $\sim$ is symmetric, reflexive and transitive, and as such, it is a equivalence relation. The geometrical interpretation for the equivalence classes may be obtained by picking up $\vec{x} \in \mathds{R}^{n+1}$ and looking for all the $\vec{y} = \lambda \vec{x}$, with real $\lambda$. They are straight lines, with $\vec{x}$ being the corresponding director vector. We will denote the quotient space by $\mathds{R}P^n := \mathds{R}^{n+1}/\sim$ and we call it projective space. We point out that given $\vec{x} \neq \vec{0}$, we may find $\lambda$ such that $\lambda \vec{x}$ is unitary. In fact, we take $\lambda = \pm 1/\Vert x \Vert$. Hence, $\mathds{R}P^n$ may be seen as the unit sphere $S^n$, with the antipodes identified. We summarize this first steps in the Figure \ref{appendix01}, with the visual example of the sphere $S^2$ immersed on $\mathds{R^3}$. The origin was removed on purpose, once it is not in any class. In light of this observation, the notation $\mathds{R}P^n = \mathds{R}^{n+1}/ \{ \vec{0} \}$ is also used. 

Complex projective spaces are defined accordingly and they are of interest to quantum mechanics \cite{bengtsson_geometry_2006}. In fact, let a qubit be represented by the vector 
\begin{equation}
    \ket{\psi} = a \ket{\alpha} + b \ket{\beta}.
\end{equation}
As usual, $\mathcal{B} = \{ \ket{\alpha}, \ket{\beta} \}$ is a basis to the space $\mathds{C}^2$ and the coefficients $a = x_1 + i x_2$, $b =x_3 + ix_4$ are complex numbers, with $x_i \in \mathds{R}$, for $i=1,2,3,4$. The normalization condition $\braket{\psi \vert \psi}=1$ implies
\begin{equation}
    x^2_1+x^2_2 + x^2_3 + x^2_4 = 1.
\end{equation}
This result allows us to conclude that the quantum state lives on the sphere $S^3 \subset \mathds{R}^4$. What happens when we change the state vector $\ket{\psi}$ by a phase factor, say, $\ket{\psi'} = e^{i \varphi}\ket{\psi}$, with $\varphi \in \mathds{R}$? They represent the same state, once probabilities are kept untouched. To see this, we associate the basis $\mathcal{B} = \{ \ket{\alpha}, \ket{\beta} \}$ to a test, with classical alternatives, say, $\alpha$ and $\beta$. The probability of finding $\alpha$ ($\beta$), according to the Born rule, is given by $\vert a \vert^2$ ($\vert b \vert^2$), when the system is in the state $\ket{\psi}$. On the other hand, the result is the same if the systems is now on the state $\ket{\psi'}$,
\begin{equation}
    Pr(\alpha \vert e^{i \varphi}\ket{\psi}) = \vert \bra{\alpha} e^{i \varphi}\ket{\psi} \vert^2 = \vert a \vert^2 = Pr (\alpha \vert \ket{\psi} ).
\end{equation}
In this case, we call the states $\ket{\psi}$ and $\ket{\psi'}$ indistinguishable. Here, $Pr(\alpha \vert \ket{\psi})$ represents the probability of finding $\alpha$ when  the system is in the state $\ket{\psi}$. 

At this stage, we are ready to connect projective complex spaces to the geometry of indistinguishable states. Unitary vectors on $\mathds{C}^2$ describe two-level systems. We denote them $(a,b)$ with the restriction $\vert a \vert^2 + \vert b \vert^2 =1$. Let us define the following relation
\begin{equation}\label{5.1}
    (a,b) \sim (c,d) \Leftrightarrow (c,d) = e^{i \varphi} (a,b), \, \varphi \in \mathds{R}.
\end{equation}
We observe the very same structure of \eqref{1.1}, which allows us to conclude that $\sim$ is indeed a equivalence relation. Just as a matter of completion, we point out that the equivalence classes in this case are orbits of the action of the group $\mathcal{U}(1)$ on $S^3$. 

Each class $[(a,b)] \subset\mathds{C}^2/ \{ (0,0) \} $ may be uniquely defined by the following map
\begin{align}\label{5.2}
    h: \mathds{C}^2/ \{ (0,0) \} &\rightarrow \mathds{C} \nonumber \\
    [(a,b)] & \mapsto h \left ( [(a,b)] \right ) = \frac{b}{a}, \, a \neq 0. 
\end{align}
Clearly $h$ is a map defined for classes, once its action is independent of the class representative. In fact, let $(c,d)$ be an arbitrary element in  $[(a,b)]$, that is,  $(c,d) = e^{i\varphi}(a,b)$. Hence, 
\begin{equation}
    h\left ( [(a,b)] \right ) = \frac{b}{a} = \frac{d}{c}.  
\end{equation}

\begin{figure}
    \centering
    \includegraphics[scale=0.32]{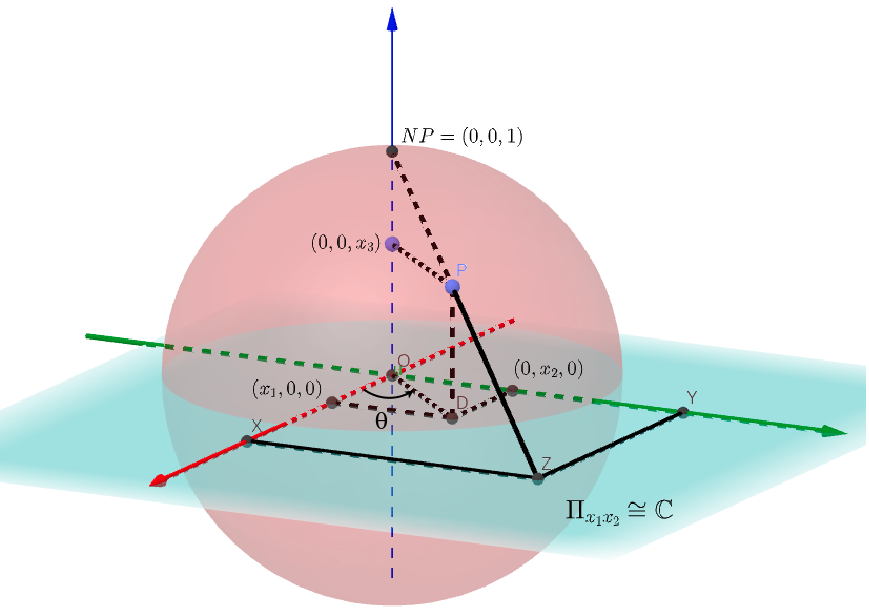}
    \caption{Schematic representation of the stereographic projection.}
    \label{6.1}
\end{figure}

A carefully look at the expression \eqref{5.2} shows that the target space of the map $h$ is $\mathds{C}$, which, in turn, may be bijectively mapped onto the sphere $S^2 \subset \mathds{R}^3$. Let us construct such bijection with what is called equatorial stereographic projection. For that, first we identify the plane $\Pi_{x_1 x_2} \subset \mathds{R}^3$ with $\mathds{C}$: $\forall Z = x_1+ix_2 \in \mathds{C},  \exists ! \, (x_1,x_2,0) \in \Pi_{x_1x_2 }$ and vice-versa. For every point $P = (x_1,x_2,x_3)$ of the sphere, except the ``north pole'' $NP$ (with coordinates $(0,0,1)$), we draw the straight line connect them, till the intersection with  $Z \in \Pi_{x_1 x_2}$, with coordinates $(X, Y, 0)$. Let us explicitly construct this map
\begin{align}
    e: S^2 \subset \mathds{R}^3 &\longrightarrow \mathds{C} \nonumber \\
    (x_1,x_2,x_3) &\longmapsto e(x_1,x_2,x_3) = X + i Y = \rho e^{i \theta}.
\end{align}
Figure \ref{6.1} helps us with the notation: $\rho = \sqrt{X^2+ Y^2}$ and $\theta = \arctan \left ( \frac{Y}{X} \right )$
Our task now consists of finding the dependence of both $\rho$ and $\theta$ in terms of the coordinates of the point $P$. Due to the similarity between the triangles $\triangle (NP, O, Z)$ and $\triangle (P, D, Z)$, we may write
\begin{equation}\label{rho.01}
    \frac{d(O,Z)}{d(O, NP)} = \frac{d(D,Z)}{d(D,P)} \leftrightarrow \rho x_3 = \rho - \sqrt{x^2_1+ x^2_2}. 
\end{equation}
Since $P \in S^2$, we have $\sqrt{x^2_1+ x^2_2} = \sqrt{1-x^2_3}$. Finally, from \eqref{rho.01}, and also noting that $\frac{Y}{X} = \frac{x_2}{x_1}$, we find
\begin{equation}\label{7.1}
    e(x_1,x_2,x_3) = \frac{\sqrt{1-x^2_3}}{1-x_3} e^{i\theta}; \quad \theta = \arctan \left (\frac{x_2}{x_1} \right ).
\end{equation}

The inverse route can also be constructed, that is, 
\begin{align}
    e^{-1}: \mathds{C} &\rightarrow S^2  \nonumber \\
    z = X +iY &\mapsto e^{-1}(z) = (x_1, x_2, x_3). 
\end{align}
This map will be important in a while. For now, remembering that $\rho = \sqrt{X^2+Y^2}$ and, together with \eqref{7.1}, we can find $x_3$ as a function of $X$ and $Y$,
\begin{equation}\label{8.1}
    \rho = \frac{\sqrt{1-x^2_3}}{1-x_3} = \sqrt{X^2+Y^2} \Rightarrow x_3 = \frac{X^2+Y^2-1}{X^2+Y^2+1}.
\end{equation}
Returning once again to the Figure \ref{6.1}, we observe that 
\begin{eqnarray}\label{8.2}
    \tan \theta = \frac{Y}{X} = \frac{x_2}{x_1} \Rightarrow x_2 = \frac{Y}{X}x_1.
\end{eqnarray}
Now we combine the constraint $x^2_1+ x^2_2+ x^2_3 =1$ together with both \eqref{8.1} and \eqref{8.2},
\begin{equation}
    x^2_1+\frac{Y^2}{X^2}x^2_1+\left (  \frac{X^2+Y^2-1}{X^2+Y^2+1} \right )^2 = 1 \Rightarrow x_1 = \frac{2X}{X^2 + Y^2 +1}. 
\end{equation}
This last result back in \eqref{8.2} provides $x_2$ as a function of $X$ and $Y$,
\begin{equation}
    x_2 = \frac{2Y}{X^2 + Y^2 +1}.
\end{equation}

We are now in position to write the manifest form of the map $e^{-1}(\cdot)$,
\begin{equation}
    e^{-1}(z = X+iY) = \left ( \frac{2X}{X^2 + Y^2 +1}, \frac{2Y}{X^2 + Y^2 +1}, \frac{X^2+Y^2-1}{X^2+Y^2+1} \right ).
\end{equation}
A direct computation shows that $\Vert e^{-1}(z)\Vert^2 = 1$, after all we are mapping $\mathds{C}$ onto a sphere. 

Summing up what we have found so far, there are two maps of interest, namely, 
\begin{equation}\label{9.2}
    h: S^3\subset \mathds{R}^4 \rightarrow \mathds{C} 
\end{equation}
\begin{equation}\label{9.3}
    e: S^2 \subset \mathds{R}^3 \rightarrow \mathds{C} \Leftrightarrow e^{-1}: \mathds{C} \rightarrow S^2. 
\end{equation}
It is suggestive to bind them according to the following composition
\begin{equation}
    e^{-1} \circ h: S^3 \rightarrow S^2,
\end{equation}
which projects points of the unity three-dimensional sphere (where the qubits live on) onto the two-dimensional sphere, immersed in our physical space. Let us find $e^{-1}$ manifestly. To sustain the same notation previously used, we write
\begin{equation}\label{72}
    h([(a,b)]) = \frac{b}{a} = \frac{x_3+ ix_4}{x_1+ix_2}=\frac{ba^*}{\vert a \vert^2}; \,\, x_i \in \mathds{R}, \, i \in \{1,2,3,4\}. 
\end{equation}
    To condensate the expression above, we set $\frac{ba^*}{\vert a \vert^2} = u +iv$, where 
\begin{equation}\label{10.3}
    u = \frac{x_1 x_3 + x_2 x_4}{x^2_1+ x^2_2}, \quad v = \frac{x_1 x_4 - x_2 x_3}{x^2_1+ x^2_2}.
\end{equation}
Now, we apply $e^{-1}$ to the result in \eqref{72}, to find
\begin{equation}
    e^{-1}(u+iv) = \left ( \frac{2u}{u^2+ v^2 + 1}, \frac{2v}{u^2+ v^2 + 1}, \frac{u^2+ v^2 - 1}{u^2+ v^2 + 1} \right ).
\end{equation}
With the notation in \eqref{10.3}, obtain the identities
\begin{equation}
    u^2+v^2=\frac{\vert b \vert^2}{\vert a \vert^2}, \,\, u^2+v^2+1=\frac{1}{\vert a \vert^2}, \,\, u^2+v^2-1=\frac{\vert b \vert^2 - \vert a \vert^2}{\vert a \vert^2}. 
\end{equation}
That way, 
\begin{equation}\label{76}
    (e^{-1}\circ h)\left ([(a,b)] \right ) = \left ( 2 \mbox{Re}(b a^*), 2 \mbox{Im}(b a^*), \vert b \vert^2 - \vert a \vert^2 \right ).  
\end{equation}

We finish our exposition with a final comment. Let us interpret the map $e^{-1}\circ h$ as a projection $\pi : S^3 \rightarrow S^2$. Actually, we may call it a fibration, with $S^2$ being the base space. Different points in $S^3$ that are connected by rotation represent the same state of a qubit. This exactly what is written in \eqref{5.1}. Due to the very structure of \eqref{76}, the phase factor makes no difference in the resultant projection: $\pi(a,b) = \pi(c,d)$. Conversely, the inverse image $\pi^{-1}(P)$ of any point $P \in S^2$ is just the entire class of indistinguishable states. As we have already seen, they are orbits of the action of $\mathcal{U}(1)$ on $S^3$, which are merely great circles. The inverse image of $\pi$ are called the fibers. Thus, we may conclude that indistinguishable states are just the fibers of $\pi$, which, are circumferences, that is, $\pi^{-1}(P) \cong S^1$. All this construction is summarized in geometric terms as  
$$ S^1 \hookrightarrow S^3 \to S^2 $$ 
and is known as the Hopf fibration \cite{bengtsson_geometry_2006}.


\begin{thebibliography}{10}
	
	\bibitem{stern_weg_1921}
	O.~Stern, ``Ein {Weg} zur experimentellen {Prüfung} der {Richtungsquantelung}
	im {Magnetfeld},'' {\em Zeitschrift für Physik} \textbf{7}, 249-253 (1921).
	
	\bibitem{gerlach_experimentelle_1921_1}
	W.~Gerlach and O.~Stern, ``Der experimentelle {Nachweis} des magnetischen
	{Moments} des {Silberatoms},'' {\em Zeitschrift für Physik} \textbf{8},
	110-111 (1921).
	
	\bibitem{gerlach_experimentelle_1922_2}
	W.~Gerlach and O.~Stern, ``Der experimentelle {Nachweis} der
	{Richtungsquantelung} im {Magnetfeld},'' {\em Zeitschrift für Physik}
	\textbf{9}, 349-352 (1922).
	
	\bibitem{gerlach_magnetische_1922}
	W.~Gerlach and O.~Stern, ``Das magnetische {Moment} des {Silberatoms},'' {\em
		Zeitschrift für Physik} \textbf{9}, 353-355  (1922).
	
	\bibitem{gomes_o_2011}
	G.~G. Gomes and M.~Pietrocola, ``O experimento de {Stern}-{Gerlach} e o spin do
	elétron: um exemplo de quasi-história,'' {\em Revista Brasileira de Ensino
		de Física} \textbf{33}, 2604 (2011).
	
	\bibitem{wu_sterngerlach_2019}
	T.-Y. Wu, A.~Kumar, F.~Giraldo, and D.~S. Weiss, ``Stern–{Gerlach} detection
	of neutral-atom qubits in a state-dependent optical lattice,'' {\em Nature
		Physics} \textbf{15}, 538-542 (2019).
	
	\bibitem{cohen-tannoudji_quantum_2020}
	C.~Cohen-Tannoudji, B.~Diu, and F.~Laloë, {\em Quantum mechanics. {Volume} 1:
		{Basic} concepts, tools, and applications}.
	\newblock Weinheim: Wiley-VCH Verlag GmbH \& Co. KGaA, second edition~ed.,
	2020.
	
	\bibitem{sakurai_modern_2017}
	J.~J. Sakurai and J.~Napolitano, {\em Modern quantum mechanics}.
	\newblock Cambridge: Cambridge University Press, second edition~ed., 2017.
	
	
	\bibitem{gallego_device-independent_2010}
	R.~Gallego, N.~Brunner, C.~Hadley, and A.~Acín, ``Device-{Independent} {Tests}
	of {Classical} and {Quantum} {Dimensions},'' {\em Physical Review Letters}
	\textbf{105}, 230501 (2010).
	
	\bibitem{GoisEtAl21}
	C.~de~Gois, G.~Moreno, R.~Nery, S.~Brito, R.~Chaves, and R.~Rabelo, ``General
	method for classicality certification in the prepare and measure scenario,''
	{\em PRX Quantum} \textbf{2}, 030311 (2021).
	
	\bibitem{nielsen.chuang.2011}
	M.~A. Nielsen and I.~L. Chuang, {\em Quantum Computation and Quantum
		Information: 10th Anniversary Edition}.
	\newblock USA: Cambridge University Press, 10th~ed., 2011.
	
	\bibitem{PhysRevLett.110.150501}
	N.~Brunner, M.~Navascu\'es, and T.~V\'ertesi, ``Dimension witnesses and quantum
	state discrimination,'' {\em Phys. Rev. Lett.} \textbf{110}, 150501 (2013).
	
	\bibitem{brunner_bell_2014}
	N.~Brunner, D.~Cavalcanti, S.~Pironio, V.~Scarani, and S.~Wehner, ``Bell
	nonlocality,'' {\em Reviews of Modern Physics} \textbf{86}, 419-478 (2014).

 \bibitem{corinne} Corinne Manogue, AIP Conference Proceedings \textbf{1413}, 55 (2012).

\bibitem{nathan} Nathan L. Harshman, American Journal of Physics \textbf{87}, 237 (2019).

\bibitem{teleporting} Charles H. Bennett, Gilles Brassard, Claude Crépeau, Richard Jozsa, Asher Peres, William K Wootters, Physical Review Letters 
 \textbf{70}, 1895–1899 (1993).

\bibitem{bb84} C. H. Bennett and G. Brassard, Proceedings of IEEE International Conference on Computers, Systems and Signal Processing \textbf{175}, 8 (1984)


 \bibitem{DA18}
	C.~Duarte and B.~Amaral, ``Resource theory of contextuality for arbitrary
	prepare-and-measure experiments,'' {\em Journal of Mathematical Physics} \textbf{59}, 062202 (2018).
	
	\bibitem{ldb2019}
	L.~M. Gaio, D.~R.~T. de~Barros, and B.~F. Rizzuti, ``Grandezas f\'isicas
	multidimensionais,'' {\em Rev. Bras. Ensino F\'isica} \textbf{41}, e20180295 (2019).
	
	\bibitem{jauch_foundations_1968}
	J.~M. Jauch, {\em Foundations of {Quantum} {Mechanics}}.
	\newblock Reading, Massachussets: Addison-Wesley Pub. Co, 1968.
	
	\bibitem{barbara.2011.book}
	B.~Amaral, A.~T. Baraviera, and M.~O.~T. Cunha, {\em Mecânica Quântica para
		Matemáticos em Formação}.
	\newblock IMPA, 2011.
	
	\bibitem{GT09}
	O.~Guhne and G.~Toth, ``Entanglement detection,'' {\em Physics Reports},
	 \textbf{474}, 1-75 (2009).
	
	\bibitem{CramerEtAl10}
	M.~Cramer, M.~B. Plenio, S.~T. Flammia, R.~Somma, D.~Gross, S.~D. Bartlett,
	O.~Landon-Cardinal, D.~Poulin, and Y.-K. Liu, ``Efficient quantum state
	tomography,'' {\em Nature Communications} \textbf{1}, 149 (2010).
	
	\bibitem{Hradil97}
	Z.~Hradil, ``Quantum-state estimation,'' {\em Physical Review A}, \textbf{55},
	R1561-R1564 (1997).
	
	\bibitem{LJ15}
	D.~Jennings and M.~Leifer, ``No return to classical reality,'' {\em
		Contemporary Physics} \textbf{57},60-82 (2015).

  \bibitem{isham} Chris J Isham, \textit{Lectures on Quantum Theory: Mathematical and Structural Foundations}. London: Imperial College Press, 1995. 
	
	\bibitem{CBRM16}
	Y.~Cai, J.-D. Bancal, J.~Romero, and V.~Scarani, ``A new device-independent
	dimension witness and its experimental implementation,'' {\em Journal of
		Physics A: Mathematical and Theoretical} \textbf{49}, 305301 (2016).
	
	\bibitem{KV17}
	K.~F. P\'al and T.~V\'ertesi, ``Family of bell inequalities violated by
	higher-dimensional bound entangled states,'' {\em Phys. Rev. A} \textbf{96},
	022123 (2017).
	
	\bibitem{Brunner2008}
	N.~Brunner, S.~Pironio, A.~Acin, N.~Gisin, A.~A. M\'ethot, and V.~Scarani,
	``Testing the dimension of hilbert spaces,'' {\em Phys. Rev. Lett.} \textbf{100}, 210503 (2008).
	
	\bibitem{watrous_theory_2018}
	J.~Watrous, {\em The theory of quantum information}.
	\newblock Cambridge, United Kingdom: Cambridge University Press, 2018.
	
	\bibitem{PhysRevResearch.2.013112}
	A.~J.~P. Garner, M.~Krumm, and M.~P. M\"uller, ``Semi-device-independent
	information processing with spatiotemporal degrees of freedom,'' {\em Phys.
		Rev. Research} \textbf{2}, 013112 (2020).
	
	\bibitem{andre.rafa.impa.2021}
	A.~S. de~Carvalho and R.~M. Siejakowski, {\em Topologia e geometria de
		3-variedades, uma agrad\'avel introdu\c{c}\~ao}.
	\newblock Rio de Janeiro, Brasil: Editora do IMPA, 2021.
	
	\bibitem{bengtsson_geometry_2006}
	I.~Bengtsson and K.~{\. Z}yczkowski, {\em Geometry of quantum states: an
		introduction to quantum entanglement}.
	\newblock Cambridge: Cambridge University Press, 2006.
	
\end{thebibliography}


\end{document}